\documentstyle[12pt,aaspp4]{article}

\begin{document}

\title{Global Seismic Oscillations in Soft Gamma Repeaters}
\author{Robert C. Duncan}
\affil{Department of Astronomy \& McDonald Observatory, 
The University of Texas, RLM 15.308, Austin, Texas 78712}
\authoremail{duncan@astro.as.utexas.edu}

\bigskip
\qquad\qquad\qquad\qquad Accepted for {\it Astrophysical Journal Letters.}
\bigskip

\begin{abstract}
 There is evidence that soft gamma repeaters (SGRs) are neutron stars which
experience frequent starquakes, possibly driven by an evolving,
ultra-strong magnetic field.  The empirical power-law distribution of 
SGR burst energies, analogous to the Gutenberg-Richter law for earthquakes,
exhibits a turn-over at high energies consistent with a {\it global limit 
on the crust fracture size.}   With such large starquakes occurring, the 
significant excitation of global seismic oscillations (GSOs) seems likely.  
Moreover, GSOs may be self-exciting in a stellar crust that is strained
by many, randomly-oriented stresses.  We explain why low-order toroidal
modes, which preserve the shape of the star and have observable 
frequencies as low as $\sim 30$ Hz, may be especially susceptible to 
excitation.  We estimate the eigenfrequencies as a function 
of stellar mass and radius, and their magnetic and rotational 
shiftings/splittings.  We also describes ways in which these modes might 
be detected and damped.  There is marginal evidence for 23 ms oscillations
in the hard initial pulse of the 1979 March 5th event.  This could be due
to the $_3t_0$ mode in a neutron star with $B\sim 10^{14}$ G or less; or 
it could be the fundamental toroidal mode if the field in the deep crust of
SGR 0526-66 is $\sim 4 \times 10^{15}$ G, in agreement with other evidence. 
If confirmed, GSOs would give corroborating evidence for 
crust-fracturing magnetic fields in SGRs: $B \gtrsim 10^{14}$ G. 

\end{abstract}


\vfill
\eject
\section{Introduction}

Global shear modes in the Earth were first detected following the devastating 
Chilean earthquake of May 1960 \markcite{ben61} (Benioff, Press \& Smith
1961).  The fundamental toroidal mode, which for $m = 0$ involves a simple, 
alternating, periodic twisting of whole hemispheres in opposite 
directions, has a period of 43 minutes in the Earth.  
In general, the toroidal modes are pure shear deformations,
divergence-free, with no radial components.  They are referred to 
by the notation $_{\ell}t_n$, where the overtone number $n$
is the number of radial nodes in the eigenfunctions
(e.g.~\markcite{mcd88}McDermott, Van Horn \& Hansen 1988; hereafter MVHH).
The fundamental mode is $_2t_0$, because $\ell = 1$ 
oscillations would violate angular momentum conservation. 
Toroidal modes up to $\ell = 60$ have been detected in the Earth, 
with $P(_{60}t_0) = 2.3$ minutes (\markcite{lay95}Lay \& Wallace 1995).  

Any condensed object with a solid component can support shear modes. 
For example, old white dwarfs with crystalline 
cores have $P(_2t_0) \approx 12$ s \markcite{han79}(Hansen \& Van Horn 1979), 
but in this case, the solid core lies within an opaque, liquid mantle, so 
detection
is problematic.  Neutron stars, which have fractionally-thin solid 
crusts floating on mantles of superfluid, offer better prospects for
supporting observable shear modes. 
Because the crusts are thin and subject to little friction
from below, large-amplitude shear oscillations could be excited with much 
less energy than would be required in true solid bodies.  Moreover, 
substantial modal deformation occurs at, or very near, the observable surface. 

 The toroidal (or torsion) modes, which preserve the star's shape, 
may be especially easy to excite via starquakes, because the restoring force 
for these modes is {\it entirely} due to the relatively weak Coulomb forces 
of the crustal ions as given by the shear modulus: $\mu \sim (Ze)^2/a^4$, 
where $a$ is the ion spacing.  Other internal modes (e.g., spheroidal modes
in the crust) involve bulk compression
and vertical motion, which have to do work against the much stronger
degeneracy pressure of electrons in the outer crust and free neutrons
in deeper layers, as well as gravity.  This implies: 
(1) toroidal shear deformations require much less 
energy than do radial or compressional deformations of comparable amplitude; 
and (2) low-order $t$-modes also have unusually low frequencies.  Since the 
damping rate (e.g. via coupling to Alfv\'en modes in the magnetosphere
or other mechanisms; \S 3 below) generally increases steeply with 
oscillation frequency, both these circumstances favor the 
significant excitation of low-order $t$-modes.

For many---or all---of these reasons, shortly after the discovery of pulsars,
Ruderman \markcite{rud68} (1968) proposed that the radio pulsations found by
Hewish et al.~\markcite{hew68}(1968) were due to the
fundamental toroidal mode, $_2t_0$, in a neutron star.  
We now know that spindown-driven seismic activity
is possible in ordinary pulsars (e.g.,~\markcite{rud91}Ruderman 1991), 
but it has not proven sufficient to excite detectable crust vibrations.
However, there is another class of stars in which Ruderman's prescient 
proposal might more readily apply: soft gamma repeaters (SGRs).  

A variety of evidence seems to indicate that SGRs are neutron stars 
which experience strong and frequent starquakes 
(e.g., \markcite{tho95, tho96}Thompson \& Duncan 1995, 1996; hereafter TD95 
and TD96).  Comparative analyses of SGR and earthquake records 
(\markcite{che96}Cheng et al.~1996) indicate that the 
relative distribution of SGR burst energies and waiting times, and their
correlations, are consistent with crust fracture/slippage events, but 
not with any known
accretion- or nuclear-powered phenomenon.  This accords with the hypothesis 
that SGRs are neutron stars with ultra-strong magnetic fields,
$B \gg m_e^2 c^3/e \hbar = 4.4 \times 10^{13}$ G. \ 
Under this hypothesis, SGR bursts are the observable 
signature of starquakes driven by the stresses of an evolving magnetic 
field; and the exceptionally energetic 1979 March 5 event is 
identified as a magnetic flare \markcite{dun92}(Duncan \& Thompson 1992,
hereafter DT92; \markcite{pac92}Paczy\'nski 1992; TD95).  Such 
magnetically-active neutron stars, or 
{\it magnetars,} could form as a result of $\alpha$--$\Omega$ dynamos in hot, 
newborn neutron stars undergoing convective mixing 
(e.g., \markcite{kei96}Keil,  
Janka \& M\"uller 1996) if they are born rotating more quickly than the
relevant ``dynamo number" threshold (\markcite{dun92, dun96, tho93}DT92;
TD93; DT96).  It is even possible that some magnetars form {\it without} 
strong convection (e.g., following rotation-supported core bounce)
because differential rotation, which is generic for high-angular momentum
proto-neutron stars, could alone generate fields 
$\sim 10^{17} \, (P_{rot}/1$ ms$)^{-1}$ G, where $P_{rot}$ is the rotation
period (DT92; TD93; DT96).  Seven different estimates of
the field in SGR 0526-66 indicate $B \gtrsim 10^{14}$ G (\S 1 in TD95). 

We now consider the physics of global seismic\footnote{Here, {\it seismic} 
refers to waves excited in the 
crust by starquakes, in analogy to the common geophysical usage.  This should 
not be confused with {\it seismological,} which could refer to
any internal mode probing the interior.} oscillations (GSOs) in SGRs. \
We will focus on $t$-modes, although our results on excitation and damping 
could apply to other crustal modes (e.g., $s$-modes) as well.

\section{Eigenfrequencies and Strain Amplitudes}

The fundamental toroidal mode's period is roughly given by the time for
a shear wave to travel around the star: $P \sim R / V_\mu$, where
$V_\mu = (\mu / \rho)^{1/2}$ is the shear wave velocity. 
\ This crude estimate ignores vertical structure in the crust; but
this is a surprisingly good approximation because $V_\mu$ varies only weakly
with depth (\markcite{rud68}Ruderman 1968; \markcite{str91}Strohmayer et 
al.~1991). Accurate mode calculations have been done by \markcite{han80}Hansen 
\& Cioffi (1980) and MVHH. \ Separation of variables for the shear wave 
equation, subject to the $t$-mode conditions 
$u_r = 0$ and ${\bf \nabla \cdot u} = 0$, gives solutions of the form 
\begin{equation}
u_\theta = {W(r) \over \hbox{sin} \theta} \ 
{\partial Y_{\ell m}\over \partial\phi} \qquad ; \qquad  
u_\phi = - W(r) \ {\partial Y_{\ell m}\over \partial\theta}, 
\end{equation}
where  $(u_r, u_\theta, u_\phi)$ are displacement amplitudes in spherical
coordinates, and $Y_{\ell m} (\theta, \phi)$ are spherical harmonics.
For $n = 0$ modes, the radial eigenfunctions $W(r)$ vary little with
depth in the crust (MVHH).  

\placetable{tbl-1}

Results of MVHH's model calculations are given in the first and third
columns of Table 1.  Note that the neutron star radii listed here are smaller 
than generally thought realistic for their respective masses, 
because MVHH used an interior (sub-crust) equation of state that
is probably unrealistically soft.  We now seek to estimate 
of the observable eigenperiod for a canonical 
$M = 1.4 \, M_\odot$, $R = 10$ km neutron star (center column of Table 1),
as well as its dependence on $M$ and $R$. \ 
To do this, note that the structure of the crust 
depends upon $M$ and $R$ almost entirely through the action of
the surface acceleration of
gravity; hence so will the effective (depth-averaged) $V_\mu$.  The quantity
$P/R \propto V_\mu^{-1}$ (fourth row of Table 1) should therefore be a function 
of $g = GM/R^2$ (fifth row, tabulated in units of $10^{14}$ cm s$^{-1}$);
so we interpolate over $g$ to we estimate $P/R$ and $P$ (center column). 
Two further corrections are needed.  Improved calculations of $\mu$
(Strohmayer et al.~\markcite{str91}1991), including directional averaging 
(\markcite{oga90}Ogata \& Ichimaru 1990), imply that $P\propto \mu^{-1/2}$
is actually larger by 1.24 than found by MVHH, as quoted in Table 1. \  
Gravitational time dilation further increases the {\it observable} period. 
We find:  
\begin{equation}
P(_2t_0) = 33.6 \ R_{10} \ {0.87 + 0.13 M_{1.4} R_{10}^{-2} 
\over (1.71 - 0.71 M_{1.4} R_{10}^{-1})^{1/2}}
\quad \ \hbox{ms}, 
\end{equation}
where $R_{10} \equiv (R/10$ km) and $M_{1.4} \equiv (M/1.4 \, M_\odot)$.

 For higher-order modes, the period goes as 
$P(_\ell t_0) \propto [\ell (\ell +1)]^{-1/2}$  
\markcite{han80}(e.g., Hansen \& Cioffi 1980), so the predicted mode
spectrum, in milliseconds, is roughly $P \approx 34, 24, 18, 15 \ldots$ 
Each of these modes is $(2\ell + 1)$--fold degenerate in $m$.  (In \S 4 
we consider 
magnetic and rotational shifting and splitting of the eigenfrequencies.)
Overtone modes ($n>0$) have much shorter periods, of order a few milliseconds
or less (e.g., MVHH). 

The mode energy scales roughly as $E_m \propto R^4/M$ since the elastic
deformation energy per unit area of crust goes as $\sim g^{-1}$ in hydrostatic
equilibrium [eq.~(6) below].  Thus the quantity $(\tilde{E}M/R^4)$
is nearly constant in MVHH's models (bottom row of Table 1), where
$\tilde{E}\equiv E_m [W(R)/R]^{-2}$ (sixth row; tabulated in units of
$10^{48}$ ergs).  \ Interpolating over $g$ again, and
including the correction factor to $\mu$, we find 
\begin{equation}
E_m = 5 \times 10^{47} \ \left[{W(R)\over R}\right]^2 \ R_{10}^4 
\, M_{1.4}^{-1} \ (0.77 + 0.23 M_{1.4} R_{10}^{-2}) \ \ \hbox{erg} 
\end{equation}
for the $_2t_0$ mode energy.  The surface displacement amplitude for $m=0$ is 
then given by eq.~(1):
\begin{equation}
u_\phi = 440 \ E_{41}^{1/2} \, R_{10}^{-1} \, M_{1.4}^{1/2} 
\sin 2\theta \ \ \hbox{cm}, 
\end{equation}
where $E_{41} \equiv (E_m/10^{41}$ erg). [For simplicity, we omit the
fine adjustment factor $(0.77 + 0.23 M_{1.4} R_{10}^{-2})^{-1/2}$ in 
this and the following equation.]  
The amplitude of the modal strain is 
\begin{equation}
\psi_m = 0.9 \times 10^{-3} \ E_{41}^{1/2} \, R_{10}^{-2} \, M_{1.4}^{1/2} 
\ \sin^2\theta. 
\end{equation}
For $m\not= 0$, the angular distribution of strain is different 
but the amplitude is comparable. 

Equation (5) is a remarkable result, because: (1) the observed, prompt 
X-ray energy a single bright SGR burst is $\sim 10^{41}$ ergs 
\markcite{nor91}(e.g., Norris et al.~1991) so the energy of a major 
starquake must be comparable or larger than this; and 
(2) the static critical (yield) strain of a neutron star crust is probably 
in the range $\psi_{cr}\sim 10^{-3}$, and perhaps less 
\markcite{smo70, rud91}(Smoluchowski \& Welch 1970; Ruderman 1991).  
[Although $\psi_{cr}\sim 10^{-2}$ in a pure bcc Coulomb
crystal, realistic crustal material is almost certainly weakened by 
lattice dislocations.]
Thus if a seismic episode, involving multiple starquakes, imparts to the 
mode even a minor fraction of the X-ray energy of a {\it single} bright 
burst, then the modal strains themselves will be large enough to trigger 
starquakes at other sites in the crust that have pre-existing 
(magnetically-induced) strains significantly less than 
$\psi_{cr}$.  \ These triggered events 
can give impetus to the mode, as we now explain. 

\section{Mode Excitation, Damping and Detectability}

The energy of an event in which critical strains are relieved throughout
an area $A$ of the crust is
${\cal E} = (1/2) \, \psi_{cr}^2  A \, \xi \, \int \mu \, dz$, 
where the integral runs over the depth of the fracture, and  $\xi> 1$
accounts for released {\it magnetic} (as opposed
to elastic) energy. (This satisfies $\xi < (4\pi \mu/B^2) \sim 10 B_{15}^{-2}$ 
except when there is significant reconnection; \S 2.2 in TD95.)  
Using the equation of state of 
\markcite{neg73}Negle \& Vautherin (1973) to evaluate the integral down to
the base of the crust, we find 
\begin{equation}
{\cal E} = 4.2 \times 10^{38} \  l_{1}^2 \, \psi_{cr3}^2
\, \xi \, R_{10}^2 \, M_{1.4}^{-1}
\ \hbox{erg,}
\end{equation}
where $\psi_{cr3}\equiv (\psi_{cr}/10^{-3})$ and 
we take $A = l_f^2$ for a fracture of length 
$l_f = 1 \, l_{1}$ km.
This result scales as $g^{-1} \propto R^2 M^{-1}$ because $dz = (\rho g)^{-1}
dP$ in hydrostatic equilibrium.  

The empirical burst energy distribution for SGR 1806--20 is a power law
with index $\gamma = 1.66$ \markcite{che95}(Cheng et al.~1995), 
analogous to the Gutenberg-Richter law for earthquakes 
\markcite{gut56}(Gutenberg \& Richter 1956).
This is a natural consequence of self-organized criticality (e.g., 
\markcite{bak96}Bak 1996). Unlike the case of earthquakes,
the SGR burst distribution turns over 
at high ${\cal E}$, with an apparent upper limit
${\cal E}_{max} = 3 \times 10^{41} \, D_{14}^2$ ergs, for SGR 1806--20
at a distance $D=14 \, D_{14}$ kpc from Earth 
(\markcite{cor97}Corbel et al.~1997).
If this is due to a maximum fracture scale, $l_{f,max}$, then
\begin{equation}
(l_{f,max}/ 2 \pi R) = 0.4 \ 
\psi_{cr3}^{-1} \, \xi^{-1/2} \, f^{-1/2} \, R_{10}^{-2} \, M_{1.4}^{1/2} 
\, D_{14} \ ,  
\end{equation}
where $f\lesssim 1$ is the fraction of the released energy that powers 
burst emissions.
This suggests that {\it the observed turn-over in the SGR burst
energy distribution is due to global limits on the fracture size.}  For
comparison, the largest earthquakes yet measured by modern seismographs
have dimensions $(l_f / 2 \pi R_\oplus) < 10^{-2}$ 
\markcite{sch90} (Scholz 1990).   

Global fractures propagate across the crust on time scales $\sim R/V_\mu$,
comparable to the period of low-order seismic modes, which are thus readily 
excited.  Once GSOs begin, new fractures can be triggered, 
but {\it always as the modal strain is locally increasing with a component 
in the direction of the pre-existing (magnetic) strain.}  
Each of these triggered fractures can give an energy boost to the mode, for 
reasons that may be understood via a simple 1-D analogy.  Imagine that
a rubber band is stretched to near its breaking point and held.  This requires
energy, say from person {\bf A} who stretches the rubber band; thus {\bf A} 
plays a role analogous to the magnetic field.  A second person, {\bf B}, 
grasps the band just within {\bf A}'s fingers and imparts a further, 
small-amplitude periodic stretching, as done by the GSO.  \ 
When the band breaks, it snaps against {\bf B}'s
fingers, giving an impulse in the direction of {\bf B}'s motion (a 
mode-boosting impulse).  The energy released is much greater than the
energy of stretching by {\bf B} (the mode); it ultimately comes from {\bf A}
(the magnetic field) as stored in the elastic band (the crustal strain).
After the band breaks, person {\bf A} might also do additional work in
dragging {\bf B}'s fingers apart.  This is the effect quantified by 
$\xi$ above.  It depends upon the fault slippage, which is difficult
to estimate (\S 2.2 in TD95).

  This analogy suggests that {\it GSOs are potentially
self-exciting in a stellar crust with many randomly-oriented strains.}  
Indeed, for a power-law distribution of pre-existing strains,
 $[\partial^2 {\cal N}/ \partial l \, \partial\psi] \propto
l^{-\sigma} \ \psi_o^{\alpha}$,
and for an excitation efficiency $\epsilon = \epsilon_o \, l_1^\beta$,
self-sustaining mode growth occurs to modal strain amplitudes 
$\psi_m \gtrsim (\varphi_m \cdot \psi_{cr})$ \ if \
$\epsilon_o > \chi \, \varphi_m \, (1+\varphi_m)^{-1} \, 
(1-\varphi_m)^{-(\alpha+1)}$, where
\begin{equation}
\chi = 0.47 \cdot (27)^{-\beta} \ 
\left({3+\beta - \sigma \over (\alpha +1)(3-\sigma)}\right) \
\psi_{cr3}^\beta \, \Lambda^{-1} \, \xi^{(\beta/2)-1} \, 
R_{10}^{\beta+2} \, M_{1.4}^{-\beta/2} \ ,
\end{equation}
and where $\Lambda$ is the fraction of the star's surface area that is
strained.
The SGR ``Gutenberg-Richter Law" (Cheng et al.~1995) implies $\sigma = 2.32$
if most fractures occur when $\psi_m = \psi_{cr} - \psi_o \ll \psi_{cr}$.
\ As long as $\beta > \sigma - 3 \approx -0.68$, large fractures dominate
the excitation.  This is probably true since large fractures tend to be 
more efficient at excitation than small, i.e., $\beta > 0$. \  
For $\beta = 1 = \alpha$ and $\varphi_m \ll 1$, the efficiency criterion is 
$\epsilon_o \gtrsim 0.002 \, (\varphi_m/0.1) \, \ 
\psi_{cr3} \, \Lambda^{-1} \, \xi^{-1/2} \, R_{10}^3 \, M_{1.4}^{-1/2}$. \
This suggests that terminal mode amplitudes 
$\sim \psi_{cr}$ could be attained by self-sustained mode growth. 

When excitation is dominated by large fractures, it is possible that only
a few low-order modes are substantially excited.  For high-$(\ell,m)$ modes, 
a randomly-oriented fault line can traverse modal zones with 
changing senses of strain, which tends to hamper excitation.  
As starquakes are triggered, the mode axis will wander (since excitation 
amplitudes add in phase like axial vectors) allowing strains to be released 
all over the crust. Thus modes which happen to be excited by a 
spontaneous starquake at the start of a seismic episode could grow
rapidly, leaving little elastic/magnetic energy to be tapped by other modes.   

How might GSOs be detected?  \ SGR burst emissions could be significantly
modulated at the mode period.  For example, a large crust fracture, once
began, could extend in length each time the mode stresses it the right
way, producing new pulses of Alfv\'en waves, pair creation, and observable
X-rays during each during each GSO cycle (cf.~\S2 in TD95).   Alternatively, 
there may be detectable discreteness in the time-spacings of the 
aftershock-like events following a bright burst 
(\markcite{kou96}Kouveliotou et al.~1996), if these are triggered by the
mode rather than by the general redistribution of stress.  
Since triggering usually occurs 
near but just before the two instants of maximum deformation in the 
mode cycle, {\it the expected time intervals between mode-triggered bursts 
are $\Delta t \simeq \eta (P/2)$, where $\eta$ is an
integer.}   This is especially true for 
$\Delta t \gtrsim P$.  \ When $\Delta t \gg P$, especially
over intervals containing bursts, modal phase drift would mask the effect.  

How do GSOs damp? \ There are several possible mechanisms; here we will
concentrate on {\it fault line reslippage and refracturing}.
Like the Earth, a neutron star is probably laced with fault lines,
along which the effective critical strain is smaller than usual: 
$\psi_{cr, f} < \psi_{cr}$.  These may be sites of recent fractures
that have not fully healed (in which case $\psi_{cr,f}$ is actually the
threshold for slipping against friction rather than a true yield
strain; Sholtz 1990), or areas where
magnetic stresses have a history of accumulating and weakening the crust. 
Whenever the modal strain amplitude along the fault line, 
$\psi_{m,f}$, exceeds $\psi_{cr,f}$, then
slippage occurs, as often as $\sim 2 \, \psi_{m,f}/ \psi_{cr,f}$ 
times per cycle when this number exceeds two.  This ``sticky" motion
reprocesses mode energy, mostly into seismic 
waves of peak frequency $\sim V_\mu/l_s$, which is in the
$\sim$ kilohertz range for $l_s \sim 1$ km, comparable
to the depth of the crust.  Such high-frequency waves couple strongly to
Alfv\'en modes in the lower magnetosphere (TD95; \markcite{tho97}Thompson 
\& Blaes 1997), while low-frequency ones are internally reflected 
(\markcite{bla89}Blaes et al.~1989; paper in preparation).  Each 
reslippage/refracture event thus injects 
energy into the magnetosphere, maintaining an X-ray emitting plasma there. 
The resultant mode damping time $\tau_{s}$, is difficult to estimate, but if 
$\tau_{s} <  130 \, E_{41} \, F_7^{-1} \, (D/8 \,\hbox{kpc})^{-2}$ s, where 
$F_{th} = 10^{-7} \, F_7$ erg cm$^{-2}$ s$^{-1}$ is
the X-ray detector threshold, then the energy of GSO damping would be
detectable as quasi-steady X-rays during active episodes, fading after
bursts cease.
Such emissions could be detectable even if many different seismic modes
are excited.  

\section{Magnetic Frequency Shifts and the March 5th Event}

If a tangled magnetic field of r.m.s. strength $B$ is embedded 
throughout in the crust, then magnetic tension augments $\mu$, and
\begin{equation}
P \approx P_o \, [ 1 + (B / B_\mu)^2 ]^{-1/2},
\end{equation}
where $P_o$ is the non-magnetic period (eq.~[2], in the case of
$_2t_0$), and $B_\mu \equiv (4 \pi \mu)^{1/2}\approx 4 \times 10^{15}
\, \rho_{14}^{0.4}$ G,  
%at $\rho_{14} \equiv (\rho / 10^{14}$ gm cm$^{-3}) = 1$, 
based on a power-law fit to the deep-crust 
($\rho > \rho_{drip}= 4.6 \times 10^{11}$ gm cm$^{-3}$) equilibrium
composition found by \markcite{neg73}Negle \& Vautherin (1973). 
Note that the mode energy (eq.~3) also shifts, upward by a factor 
$[1+(B/B_\mu)^2]$.

 Even if $B\ll B_\mu$ in the {\it deep} crust, near the star's surface, 
at densities
$\rho < 4\times 10^{10} \, B_{14}^{1.54}$ gm cm$^{-3}$ for $B_{14}< 4$,
magnetic restoring forces exceed elastic ones.
This perturbs the eigenfunctions in the outer
layers of the crust, but it has little effect on the observable GSO period 
since the great preponderance of mode deformation energy (or kinetic energy) 
resides below this \markcite{car86}(Carroll et al.~1986). 
Thus it is appropriate to take $\rho_{14}\sim 1$ in the formula for 
$B_\mu$ when evaluating eq.~(9) [except in the Alfv\'en limit,
$B\gg B_\mu$, for which eq.~(9) no longer applies]. 

 Magnetic frequency shifts also can {\it split} the eigenfrequencies,
resolving the $(2\ell +1)$--fold degeneracy in $m$. \  Equation (9)
is precisely accurate only in the case of a disordered field uniformly 
penetrating the star's crust.  A realistic field with 
global structure  will produce
$m$-dependent period shifts of order $(\Delta P /P) \sim (B/B_\mu)^2$ for 
$B\ll B_\mu$, where the precise splittings depend upon 
field geometry and mode orientation. 
Rotational splittings $\Delta P/P = [m/\ell (\ell +1)] \, (P /P_{rot})$
also occur (\markcite{lap81}Lapwood \& Usami 1981; 
\markcite{str91}Strohmayer 1991).  However, for $_2t_0$, this is only 
$\Delta P/P = 7\times 10^{-4} \, m \, (P_{rot}/ 8.0 \, \hbox{s})^{-1}$,
where we scale to the rotation period of SGR 0526--66 as inferred from 
the 1979 March 5th event. For comparison, rotational splittings 
of $t$-modes are $\sim 10$ times larger in the Earth, yet they have not 
been detected (\markcite{lay95}Lay \& Wallace 1995).

The exceptionally energetic 1979 March 5th event was probably due to a
large-scale instability in a magnetar (\S 2.1 of TD95).  
A likely candidate is the Flowers-Ruderman instability, which involves a
global crust fracture and a (small-angle) relative twist of two sections
of crust, diminishing the exterior dipole moment as reconnection powers 
the flare (\markcite{flo77}Flowers \& Ruderman 1977; 
\S 14.2 and \S 15.2 in TD93).  This 
would excite very strong GSOs.  Indeed, there is evidence for a 23--ms 
periodicity in the hard initial pulse of the event 
(\markcite{bar83}Barat et al.~1983).  
This could be due to $_3t_0$ oscillations in a neutron star with  
$B\lesssim 10^{14}$ G; or it could be due to the fundamental, $_2t_0$, if
the field in the deep crust is $B = 1.06 \, B_\mu \approx 4 \times 10^{15}$
G, for a $1.4\, M_\odot$, 10 km neutron star (eqs.~[1] and [9]).
Such strong fields in the deep crust of SGR 0526--66 have previously 
been inferred for other reasons (TD95; TD96).
Note that several modes would probably be excited by such 
a catastrophe, and the ongoing release of energy would also affect 
crustal motions, so a complex (non-sinusoidal) light curve 
is to be expected.

In conclusion, if SGR bursts are due to starquakes, then strong seismic
waves almost certainly vibrate SGR crusts during active episodes. 
Whether a few low-order GSOs can attain sufficient amplitudes to have
detectable consequences is less certain. We have given three reasons
why this is at least possible: (1) global-sized fractures (eq.~[7]) naturally
excite low-order modes; (2) mode damping, via Alfv\'en wave loss and other 
mechanisms, declines steeply with frequency; and (3) the 
``self-exciting" effect could drive runaway growth of single modes.  

If the presense of GSOs in SGRs, as suggested by the 1979 March 5th data,  
is verified by future observations, this will add
to the (already considerable) evidence for seismic activity 
in these stars.  Magnetism is a natural candidate
for driving this activity, especially in light of the other
evidence for very strong fields in SGRs (as summarized in \S 1 of TD95). \
Rotational energy may trigger some starquakes and glitches in 
ordinary pulsars (e.g., \markcite{rud91}Ruderman 1991),  but this 
could not power SGR activity in a neutron star with an 8.0 s period 
like SGR 0526--66. \ Magnetars seem necessary, because typical 
pulsar-strength fields  could not drive strains in excess of $\psi_{cr}$ 
and fracture the crust.  Hooke's Law requires 
\begin{equation}
B > ( 4 \pi \, \mu \, \psi_{cr})^{1/2} = 1.2\times 10^{14} \ \rho_{14}^{0.4}
\ \left({\psi_{cr}\over10^{-3}}\right)^{1/2} \ \ \hbox{G,}
\end{equation}
where $\rho_{14}\sim 1$ because only deep fractures have sufficient energy
to power bright SGR bursts (eqs.~[6]--[7]). 

\acknowledgments

I thank C. Thompson and C. Kouveliotou for discussions. 
This work was supported by the NASA Astrophysics Theory 
Program, grant number NAG5-2773; and by the Texas Advanced 
Research Program, grant number ARP-028.

\clearpage

\begin{deluxetable}{clll}
%\small
\tablecaption{Eigenperiod and Energy of the Fundamental Toroidal Mode
\label{tbl-1}}
\tablewidth{300pt}
\startdata
$M/M_\odot$     &0.503   &1.4     &1.326 \nl
$R$ (km)        &9.785   &10      &7.853 \nl
$P$ (ms)        &18.54   &20.7    &17.32  \nl
$P/R$ (ms/km)   &1.89    &2.07    &2.21  \nl
$g_{14}$        &0.701   &1.87    &2.87  \nl
$\tilde{E}_{48}$  &1.60  &0.73 &0.331 \nl 
$\tilde{E}_{48} M_{1.4}/ R_{10}^4$  &0.63  &0.73 &0.82 \nl

\tablecomments{Columns 1 and 3 show results for models NS05T8 and
NS13T8 of McDermott, Van Horn \& Hansen (1988), which employ a soft
interior equation of state.  Column 2 gives extrapolated values for 
a canonical model star.  Further corrections are needed before comparing
with observations, as explained in the text.}
\enddata
\end{deluxetable}

\end{document}